\documentclass[aps, prd, twocolumn, superscriptaddress, nofootinbib]{revtex4-1}

\usepackage{amsmath}	
\usepackage{amsthm}		
\usepackage{amssymb}	
\usepackage{eufrak}
\usepackage{datetime}
\usepackage{graphicx}
\usepackage{color}
\usepackage{verbatim}
\usepackage{bm}
\usepackage{float}

\usepackage[colorlinks=true, citecolor=midblue, linkcolor=midblue, urlcolor=midblue]{hyperref}

\usepackage{setspace}

\usepackage[normalem]{ulem}

\settimeformat{ampmtime}

\definecolor{grey}{rgb}{0.4,0.4,0.4}
\definecolor{dullmagenta}{rgb}{0.4,0,0.4}
\definecolor{darkblue}{rgb}{0,0,0.4}
\definecolor{midblue}{rgb}{0,0,0.5}
\definecolor{midred}{rgb}{0.5,0,0}
\definecolor{orange}{rgb}{1,0.5,0}
\definecolor{lightbrown}{rgb}{0.75,0.5,0.25}
\definecolor{tan}{cmyk}{0.14,0.42,0.56,0}
\definecolor{djunglegreen}{cmyk}{0.99,0,0.52,0}
\definecolor{lightgreen}{rgb}{0,1,0}
\definecolor{olivegreen}{cmyk}{0.64,0,0.95,0.40}
\definecolor{midgreen}{rgb}{0.0,0.675,0.0}
\definecolor{darkgreen}{rgb}{0,0.5,0}

\newcommand{\vs}{\vspace}

\renewcommand{\.}{\hspace{0.5mm}}

\newcommand{\Arm}{\ensuremath{\mathrm{A}}}
\newcommand{\Brm}{\ensuremath{\mathrm{B}}}

\newcommand{\brm}{\ensuremath{\mathrm{b}}}
\newcommand{\crm}{\ensuremath{\mathrm{c}}}
\newcommand{\drm}{\ensuremath{\mathrm{d}}}

\newcommand{\srm}{\ensuremath{\mathrm{s}}}

\newcommand{\Ocal}{\ensuremath{\mathcal{O}}}

\renewcommand{\d}{\ensuremath{\mathrm{d}}}

\newcommand{\keV}{\ensuremath{\mathrm{keV}}}
\newcommand{\MeV}{\ensuremath{\mathrm{MeV}}}

\newcommand{\eg}{e.g.}
\newcommand{\ie}{i.e.}

\newcommand{\cf}{c.f.}

\def\Msun{M_\odot}
\def\fPBH{f_{\rm PBH}}

\settimeformat{ampmtime}

\usepackage{float}

\begin{document}

\title{Lepton Flavour Asymmetries and the Mass Spectrum of Primordial Black Holes}

\author{Dietrich B{\"o}deker}
\email{bodeker@physik.uni-bielefeld.de}
\affiliation{
	Fakult{\"a}t f{\"u}r Physik,
	Universit{\"a}t Bielefeld,
	Postfach 100131,
	33501 Bielefeld,
	Germany}

\author{Florian K{\"u}hnel}
\email{Florian.Kuehnel@physik.uni-muenchen.de}
\affiliation{
	Arnold Sommerfeld Center,
	Ludwig-Maximilians-Universit{\"a}t,
	Theresienstra{\ss}e 37,
	80333 M{\"u}nchen,
	Germany}

\author{Isabel M.~Oldengott}
\email{isabel.oldengott@uv.es}
\affiliation{
	Departament de Fisica Te{\`o}rica and IFIC,
	CSIC-Universitat de Val{\`e}ncia,
	46100 Burjassot,
	Spain}

\author{Dominik J.~Schwarz}
\email{dschwarz@physik.uni-bielefeld.de}
\affiliation{
	Fakult{\"a}t f{\"u}r Physik,
	Universit{\"a}t Bielefeld,
	Postfach 100131,
	33501 Bielefeld,
	Germany}
	
\date{\formatdate{\day}{\month}{\year}, \currenttime}

\begin{abstract}
We study the influence of lepton flavour asymmetries on the formation and the mass spectrum of primordial black holes. We estimate the detectability of their mergers with LIGO/Virgo and show that the currently published gravitational wave events may actually be described by a primordial black hole spectrum from non-zero asymmetries. We suggest to use gravitational-wave astronomy as a novel tool to probe how lepton flavour asymmetric the Universe has been before the onset of neutrino oscillations.
\end{abstract}


\maketitle

\section{Introduction}
\label{sec:Introduction}

The ascent of gravitational-wave astronomy with its milestone discovery of two merging black holes by the LIGO/Virgo collaborations \cite{Abbott:2016nmj} has provided us with unparalleled insights into the physics of black holes. With $47$ confirmed events \cite{Abbott:2020gyp} we can now enter population studies of black holes, which might shed further light on their origin. Black holes with masses between a solar mass and several tens of solar masses can be formed in the core collapse of stars in the late Universe \cite{Heger:2002by}, or primordially, \ie~prior to the epoch of star formation, \cite{1967SvA....10..602Z, Carr:1974nx} (see Ref.~\cite{Carr:2020xqk} for a recent review). One scenario for their formation is the gravitational collapse of extremely rare and large overdensities during the radiation dominated epoch.

Black holes of stellar origin are thought to be limited to masses below $\sim 50\,M_{\odot}$ \cite{2010ApJ...714.1217B, Leung:2019fgj}, while higher mass holes would have to result from (hypothetical) population III stars \cite{2020arXiv200906922K} or mergers of smaller holes of stellar origin. These double (or in general multi-)merger events happen rarely; furthermore, the calculation of their rates is non-trivial. On the other hand, there is a plethora of primordial black hole (PBH) scenarios which yield mass spectra including the observable mass range of LIGO/Virgo or LISA \cite{lisa-proposal}. The thermal history of the early Universe naturally induces peaks in the PBH formation probability, as has been pointed out in Ref.~\cite{Carr:2019kxo}.

Within the Standard Model of particle physics a softening of the equation of state occurs at several instances in the thermal history of the Universe: during the electroweak transition, the quantum chromodynamics (QCD) transition, the pion/muon annihilation and the electron-positron annihilation. This predicts peaks in the PBH mass spectrum in several mass ranges, roughly corresponding to the mass contained within the Hubble volume at the respective times (see Ref.~\cite{Carr:2019kxo} for a discussion on the entire mass range, but also Refs.~\cite{Jedamzik:1996mr, Byrnes:2018clq, Clesse:2020ghq, Jedamzik:2020omx} for work focussed on the QCD transition). More precisely, this directly modifies the PBH formation threshold which yields exponential amplification in the PBH mass function at planetary scales, around a solar mass, an order of magnitude above, and at a million solar masses, corresponding to the electroweak transition, the QCD transition, the pion/muon annihilation, as well as the electron-positron annihilation, respectively, and thereby generates an extended and multimodal PBH mass spectrum (\cf~Ref.~\cite{Carr:2018poi}).

This imprint of the thermal history on the PBH mass spectrum applies regardless whether PBHs make up $100\%$ or just a tiny fraction of the dark matter. Hence, their mass spectrum can be used to probe the very early Universe, especially by means of the recently detected mergers of black holes with masses well above $50\,M_{\odot}$. Thus the hadron and charged lepton-annihilation epochs, when the Universe had a temperature  $T \sim 40\,$MeV and $T \sim 170\,$keV, can be probed by LIGO/Virgo and by LISA, respectively.

In particular, it is possible to investigate a potential lepton flavour asymmetry, which might be present at $40\,$MeV, as the oscillation of neutrinos sets in only at  $T \sim 10\,$MeV. From then on lepton flavour is no longer conserved. Reference~\cite{Vovchenko:2020crk} calculated the PBH dark-matter fraction including large lepton flavour asymmetries for PBH masses $M > \Ocal( 1 )\,\Msun$. In this work, we extend the mass range to include all masses within the interval $\big[ 10^{-3},\.10^{3} \big]\,\Msun$, and also derive the probability distribution of PBH merger detections by LIGO/Virgo.

Lepton flavour asymmetries are defined as 
\begin{equation}
    \ell_{\alpha}
        \equiv
                    \frac{ n_{\alpha} - n_{\bar{\alpha}} + n_{\nu_{\alpha}} - n_{\bar{\nu}_{\alpha}} }
                    { s }
                    \; ,
                    \quad
                    \alpha\.\in\.\{ e,\,\mu,\,\tau \}
                    \; ,
                    \label{eq:leptonasymmetry}
\end{equation}
where $n_{\alpha}$ denotes the number density of electrons, muons and tau leptons ($n_{\bar{\alpha}}$ standing for the respective antileptons), and $s$ is the entropy density. When the expansion of the Universe is adiabatic, the $\ell_{\alpha}$ are conserved between the electroweak transition at $T \simeq 160\,$GeV and  $T \simeq 10\,\MeV$. The baryon asymmetry $b$ of the Universe [defined analogously to Eq.~\eqref{eq:leptonasymmetry}] is well constrained from observations of the Cosmic Microwave Background (CMB) \cite{Aghanim:2018eyx} and primordial element abundances \cite{Pitrou:2018cgg} and known to be a tiny number, \ie~$b = 8.7 \times 10^{-11}$ (inferred from Ref.~\cite{Aghanim:2018eyx}). Constraints on the lepton asymmetries are however many orders of magnitude weaker and allow for a total lepton asymmetry as large as~\cite{Oldengott:2017tzj}
\begin{equation}
    | \ell_{e} + \ell_{\mu} + \ell_{\tau} |
        <
                    1.2 \times 10^{-2}
                    \, .
                    \label{eq:ell-total}
\end{equation}
In particular, as pointed out in Refs.~\cite{Stuke:2011wz, Middeldorf-Wygas:2020glx, Vovchenko:2020crk}, scenarios of unequal lepton flavour asymmetries (before the onset of neutrino oscillations) are observationally almost unconstrained and therefore open up a whole new parameter space for studying the evolution of the Universe at temperatures around the QCD transition at $160 \simeq {\rm MeV}$.

\section{Thermal History of the Universe}
\label{sec:Thermal-History-of-the-Universe}

Shortly after reheating, the Universe is filled with radiation, whose energy density $\epsilon \propto g_{*}( T )\,T^{4}$ decreases as the Universe expands and the number of relativistic degrees of freedom $g_{*}$ successively lowers as more and more particles become non-relativistic. In the Standard Model, the first particle to become non-relativistic is the top quark, followed by the Higgs, the $Z$ and $W$ bosons. Below $T \sim 5\,$GeV, the bottom and charm quarks, as well as the tau disappear from the plasma before the largest change in $g_{*}$ happens at the QCD transition, when all remaining quarks and the gluons bind to form mostly pions. After this, at $T \sim 40\, \MeV$, pions and muons become non-relativistic, and finally at $T \sim 170\,\keV$, $e^{+}e^{-}$ annihilation causes a further drop of $g_{*}$.

\begin{figure}
	\centering
	\includegraphics[width = 0.40 \textwidth]{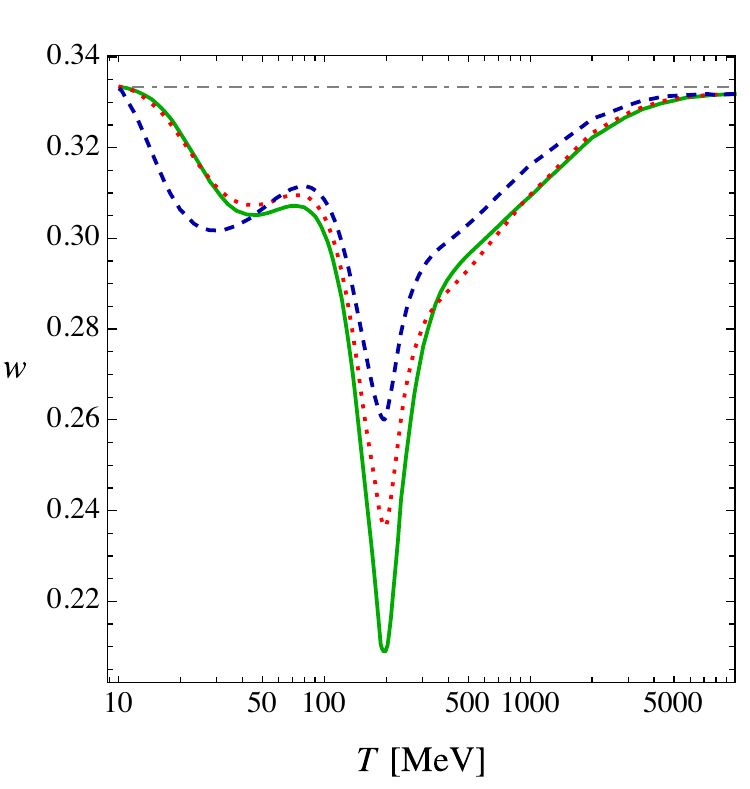}
	\caption{Effective equation-of-state parameter 
	        as a function of temperature.
	        Shown are the three cases of different 
	        lepton flavour asymmetries 
	        as discussed in the main text as well as 
	        a constant line at $1 / 3$ indicating the 
	        case of an ideal radiation fluid. 
	        The standard scenario [case ($i$)] is indicated by the green, solid line.}
			\vs{-6mm}
	\label{fig:EoS}
\end{figure}

Each of these events changes the effective equation of state parameter $w \equiv p / \epsilon$. In a series of works \cite{Stuke:2011wz, Wygas:2018otj, Middeldorf-Wygas:2020glx} it has been shown how large lepton (flavour) asymmetries induce large chemical potentials for the different particle species.  By applying the method outlined in Ref.~\cite{Middeldorf-Wygas:2020glx} in Fig.~\ref{fig:EoS} we show the effect of lepton flavour asymmetries on $w$ in the temperature range $[ 10,\,10^{4} ]\,\MeV$, the epoch around the cosmic QCD transition.\footnote{The equation of state is obtained from a smooth interpolation of three data sets (ideal quark-gluon gas, lattice QCD including 2+1+1 quark flavours, hadron resonance gas; see Ref.~\cite{Middeldorf-Wygas:2020glx} for details). We have explicitly checked that the concrete interpolation does not significantly affect the PBH abundance. Note that the effective equation of state for the standard scenario has been obtained in Ref.~\cite{Laine:2006cp}.} Concretely, we show the three cases of
\begin{itemize}

	\item[({\it i}$\mspace{1.5mu}$)] 
	    $\ell_{e} = \ell_{\mu} = \ell_{\tau} = -\,5.3 \times 10^{-11}$\\~
	    [green, solid line];

	\item[({\it ii}$\mspace{1.5mu}$)] 
	    $\ell_{e} = 0$ and $\ell_{\mu} = -\,\ell_{\tau} = 4 \times 10^{-2}$\\~
		[red, dotted line];

	\item [({\it iii}$\mspace{1.5mu}$)] 
	    $\ell_{e} = -\,8 \times 10^{-2}$ and $\ell_{\mu} = \ell_{\tau} = 4 \times 10^{-2}$\\~
		[blue, dashed line].

\end{itemize}
The first case represents the standard scenario (see \eg~Ref.~\cite{Laine:2006cp}) where the lepton asymmetry is assumed to be related to the baryon asymmetry by $\ell = -\.( 51 / 28 )\.b$ \cite{Kolb:1983ni} due to sphaleron processes.\footnote{Reference \cite{Carr:2019kxo} utilises a slightly different numerical result for the equation-of-state parameter implying corresponding deviations in the derived quantities.} However, neither has the existence of sphalerons been proven experimentally, nor did we identify the origin of the matter-antimatter asymmetry of the Universe yet. The cases ({\it ii}$\mspace{1.5mu}$) and ({\it iii}$\mspace{1.5mu}$) serve as benchmark models in order to illustrate the possible influence of lepton flavour asymmetries on PBH formation. The numerical values are chosen such that they maximise the impact on the equation of state, satisfy existing limits on the total lepton asymmetry \eqref{eq:ell-total}, and at the same time guarantee the applicability of the Taylor expansion of the equation of state as discussed in Ref.~\cite{Middeldorf-Wygas:2020glx}.

Beyond the cases shown in Fig.~\ref{fig:EoS} we also investigated the situation of equal lepton flavour asymmetries (\ie~$\ell_{e} = \ell_{\mu} = \ell_{\tau}$), saturating the bound of Eq.~\eqref{eq:ell-total}. It turns out that $w$ is very similar to the standard scenario and therefore not shown.

It can clearly be seen from Fig.~\ref{fig:EoS} how non-zero flavour asymmetries weaken the softening of $w$ during the transition, and that even the case with $\ell_{e} = 0$ yields a pronounced effect. 
The two cases of unequal lepton flavour asymmetry are chosen for illustrational purposes. Note that a lepton flavour asymmetry always weakens the softening of the equation of state during the QCD transition, as such an asymmetry adds leptons to the Universe which do not interact strongly. This is different for the smaller effect at the pion/muon plateau, for which different lepton flavour asymmetries can lead to stiffening or softening of the equation of state (\cf~the two cases of unequal flavour asymmetry), as $w$ sees pions and muons to become nonrelativistic.

\section{Primordial Black Hole Formation}
\label{sec:Primordial-Black-Hole-Formation}

\begin{figure}
	\centering
	\includegraphics[width = 0.40 \textwidth]{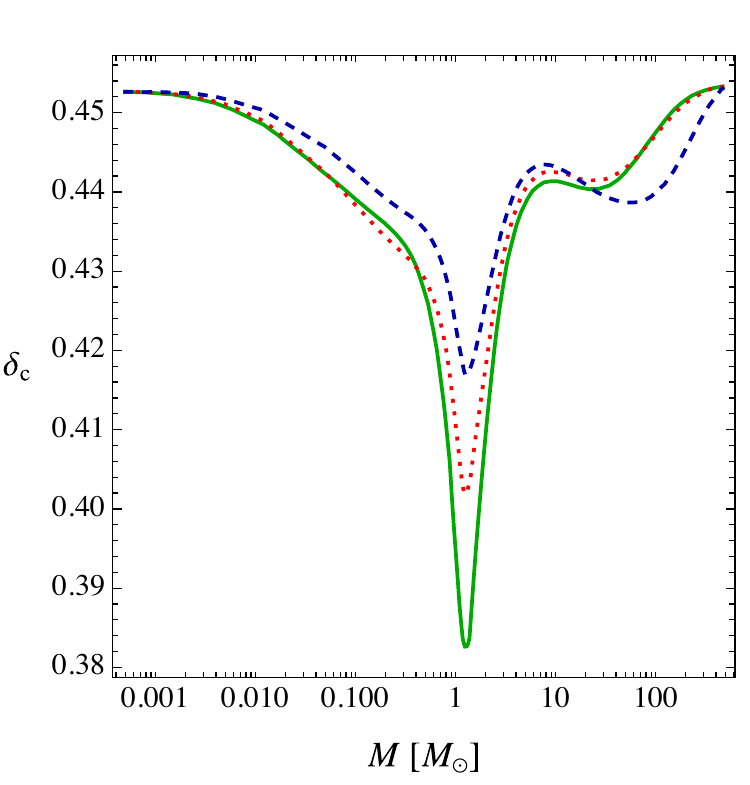}
	\caption{
			Formation threshold $\delta_{\crm}$ as a function of PBH mass $M$,
			for the three cases of different lepton flavour asymmetries as discussed in the main text. 
			The standard scenario is shown by the green, solid line.
			}
	\label{fig:deltac}
\end{figure}

PBH formation generically requires the generation of large overdensities. In most scenarios, the density contrast, $\delta \equiv \delta \epsilon / \epsilon$, is of inflationary origin. When these overdensities re-enter the Hubble horizon, they collapse if they are larger than some threshold $\delta_{\crm}$ (see Fig.~\ref{fig:deltac}), which depends on the equation-of-state parameter $w( T,\,\ell_{\alpha} )$, $\alpha \in \{ e,\,\mu,\,\tau \}$. As implied by Fig.~\ref{fig:EoS}, and studied for similar cases in Refs.~\cite{Jedamzik:1996mr, Jedamzik:1999am, 2018JCAP...08..041B, Carr:2019kxo}, the thermal history of the Universe can hence induce distinct features in the PBH mass function even for a (quasi) scale-invariant power spectrum. The underlying reason, which has been pointed out by Carr \cite{Carr:1975qj}, is that if the PBHs form from Gaussian inhomogeneities with root-mean-square amplitude $\delta_{\rm rms}$, the fraction $\beta$ of horizon patches which undergo gravitational collapse to PBHs is
\begin{equation}
	\beta( M,\,\ell_{\alpha} )
		\approx
				{\rm erfc}\!
				\left[
					\frac{\delta_{\crm}\big( w[ T( M ), \ell_{\alpha} ] \big)}
						{ \sqrt{2}\,\delta_{\rm rms}( M )}
				\right]
				.
				\label{eq:beta(T)}
\end{equation}
Here $M$ is the PBH mass and `erfc' is the complementary error function. Reference~\cite{Musco:2012au} (see the right panel of their Fig.~8) provides numerical results for the threshold $\delta_{\crm}$, which we will utilize in this work. The numerical value of the threshold $\delta_{\crm}$ depends on shape and statistics of the collapsing overdensities, \cf~Refs.~\cite{Escriva:2019phb, Musco:2020jjb} for spherical perturbations and Ref.~\cite{Kuhnel:2016exn} for non-spherical shapes. Equation~\eqref{eq:beta(T)} shows the exponential sensitivity of $\beta$ to $w$. 
The temperature at PBH formation and the mass of the corresponding PBH are related via (\cf~Ref.~\cite{Carr:2020gox})
\begin{equation}
	T
		\approx
				700\;g_{*}^{-1/4}\,\sqrt{\Msun / M\,}\;\MeV
				\, .
				\label{eq:T(M)}
\end{equation}
The present fractional PBH dark-matter spectrum is
\begin{align}
\begin{split}
    \frac{\drm\. \fPBH( M,\,\ell_{\alpha} )}{\drm \ln M} 
		\equiv
				\frac{ 1 }
					{ \rho_{\rm CDM} }
				\frac{ \drm\.\rho_{\rm PBH} ( M,\,\ell_{\alpha} ) }
					{\drm \ln M }
		        \\[2mm]
		\approx 2.4\,\beta( M,\,\ell_{\alpha} )\.
				\sqrt{
					\frac{ M_{\rm eq} }
						{ M }
				\,}
				\, ,
				\label{eq:fPBH}
\end{split}
\end{align}
where $\rho_{\rm CDM}$ and $\rho_{\rm PBH}$ denote the mass densities of cold dark matter (CDM) and PBHs, respectively, with $M_{\rm eq} \approx 2.8 \times 10^{17}\,\Msun$, being the horizon mass at matter-radiation equality. The numerical factor of $2.4$ equals $2\.( 1 + \Omega_{\brm} / \Omega_{\rm CDM} )$, with $\Omega_{\rm CDM} = 0.245$ and $\Omega_{\brm} = 0.0456$ being the CDM and baryon density parameters, respectively \cite{Aghanim:2018eyx}.

\begin{figure}
	\centering
	\includegraphics[width = 0.46 \textwidth]{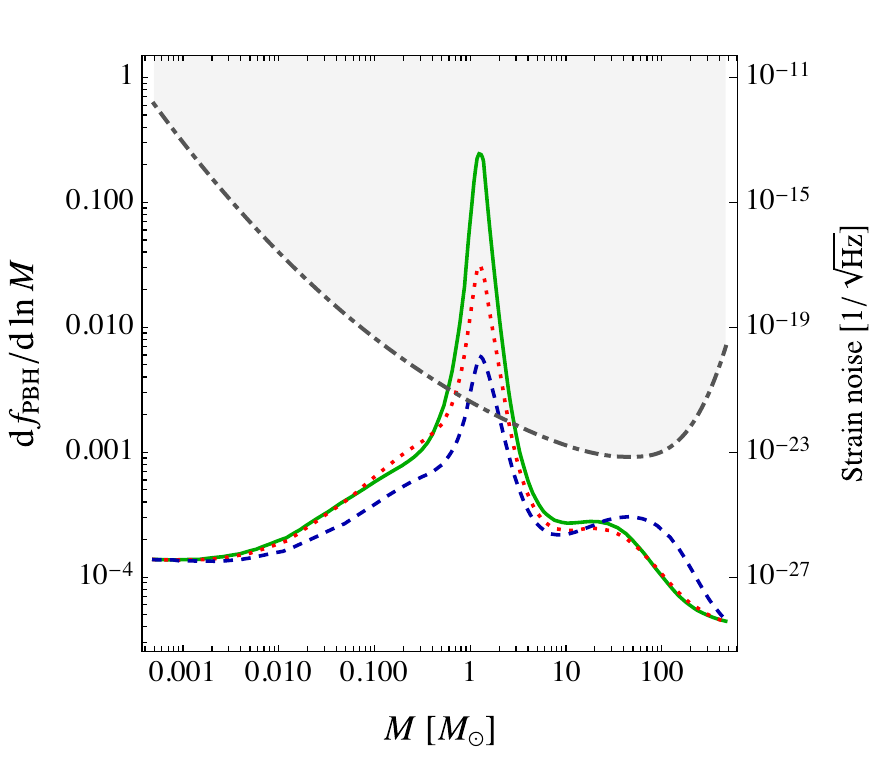}
	\caption{
			Spectral density of the PBH dark-matter fraction as a function of PBH mass $M$,
			for the three cases of different lepton flavour asymmetries as discussed in the main text.
			The green, solid line indicates the standard scenario.
			Also shown is the LIGO sensitivity curve (grey, dot-dashed line) from 
			Ref.~\cite{LIGOScientific:2018mvr}, schematically for 
			equal-mass mergers and using the maximal gravitational-wave frequency
			$f^{}_{\rm max} \approx 4400\,M_{\odot} / M$ for the conversion from frequency to mass.
			}
	\label{fig:fPBH}
\end{figure}

Following Ref.~\cite{Carr:2019kxo}, we assume a spectrum of the form\footnote{There are several ways to enhance the amplitude of the primordial power spectrum, such as in hybrid inflation \cite{Clesse:2015wea} or for inflationary potentials with a plateau feature or an inflection point \cite{Ezquiaga:2018gbw, Biagetti:2018pjj, Kuhnel:2019xes} (see Ref.~\cite{Akrami:2016vrq} for a discussion on respective constraints and uncertainties).}
\begin{equation}
	\delta_{\rm rms}( M )
		=
					(
						M / \Msun
					)^{(1 - n_{\srm}) / 4}
					\times
				\begin{cases}
					A
					& \text{[CMB scales]}\\[2.5mm]
					\tilde{A}\.
					& \text{[PBH scales]\,,}
				\end{cases}
				\label{eq:delta-power-law}
\end{equation}
wherefore the spectral index $n_{\srm}$ and amplitude $A$ are taken to assume their CMB values \cite{Aghanim:2018eyx}, $n_{\srm} = 0.97$ and $A = 4 \times 10^{-5}$, respectively. Exemplary, we adjust the amplitude $\tilde{A}$ such that the standard scenario case ({\it i}$\mspace{1.5mu})$ yields a dark-matter fraction of $10\.\%$ with the interval $[ 10,\,10^{4} ]\,\MeV$. By virtue of Eq.~\eqref{eq:T(M)}, these boundary values correspond to PBH masses of $5 \times 10^{-4}\,\Msun$ and $1500\,\Msun$, which includes the entire LIGO/Virgo sensitivity range. Demanding
\begin{align}
	0.1
	    &=
	f^{(i)}_{\rm PBH}
	    \equiv
	                \int_{5 \times 10^{-4}\mspace{1mu}\Msun}^{1500\mspace{1mu}\Msun}\!\!\!\d\mspace{-1mu}\ln M\;
		            \frac{ \d f_{\rm PBH}( M,\,\ell_{\alpha} = 0 ) }
	                { \d\mspace{-1mu}\ln M }
					\label{eq:fPBHinInterval}
\end{align}
implies $\tilde{A} \approx 0.0189$. For the unequal flavor asymmetric cases
({\it ii}$\mspace{1.5mu}$) and ({\it iii}$\mspace{1.5mu}$) this yields the integrated dark-matter fractions $f^{(ii)}_{\rm PBH} = 0.0197$ and $f^{(iii)}_{\rm PBH} = 0.0073$.

As regards the merger-rate calculation discussed below, PBH formation outside of this range won't have any significant effect. Hence, we will be agnostic on possible alterations of the equation-of-state parameter for PBH masses below $5 \times 10^{-4}\,\Msun$ and above $1500\,\Msun$. The result for the PBH dark-matter fraction is depicted in Fig.~\ref{fig:fPBH}. It is seen that, due to the exponential enhancement [see Eq.~\eqref{eq:beta(T)}], all three cases differ significantly. Below we will see how this affects the PBH merger rates. We have also studied flavour-equilibrated universes with lepton asymmetry as large as allowed by Eq.~\eqref{eq:ell-total} that respect the CMB and Big Bang Nucleosynthesis bounds, and found that these are indistinguishable from the standard case with regard to its PBH mass spectrum.

\begin{figure}
	\centering
	\includegraphics[width = 0.45 \textwidth]{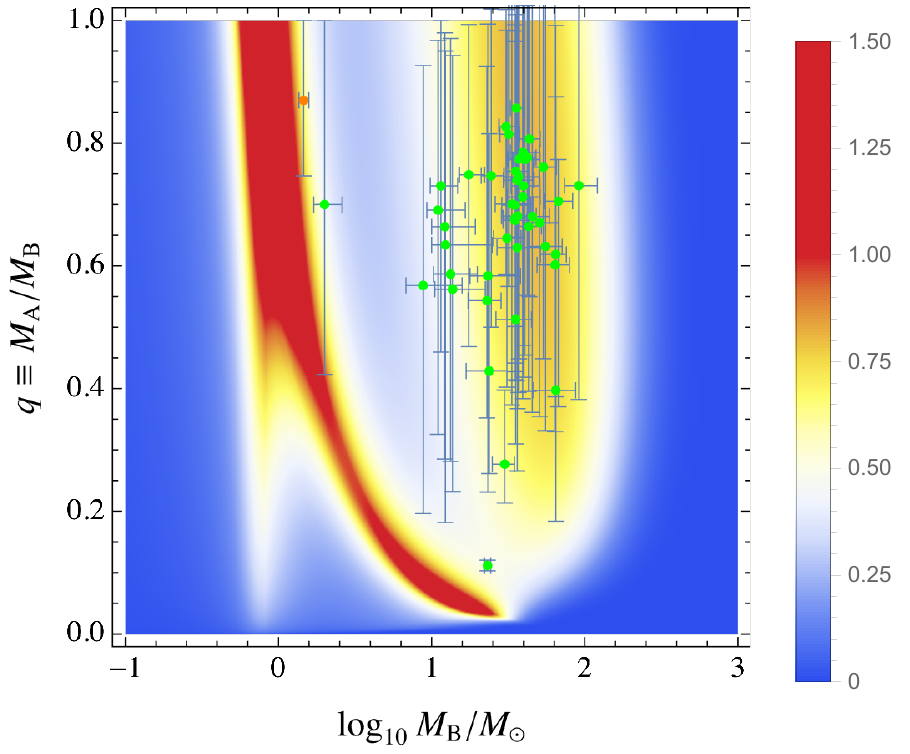}
	\vs{-1.5mm}
	\caption{Probability density
	        of PBH merger detections with masses 
			$M_{\Brm} > M_{\Arm}$ by LIGO/Virgo 
			for case ({\it i}$\mspace{1.5mu}$)
			$[ \d\.{\rm Prob.} / \d\mspace{-0.5mu}\ln( M_{\Arm} )\.\d\mspace{-0.5mu}\ln( M_{\Brm} ) ]$.
			Also shown are all currently known merger events (green dots) 
			as reported by the LIGO/Virgo collaborations \cite{Abbott:2020gyp}, including 
			the binary neutron-star merger GW170817 (orange dot).
			The parameter space is spanned by $M_{\Brm}$ and 
			the mass ratio $q \equiv M_{\Arm} / M_{\Brm}$.
			\vs{-1mm}
			}
	\label{fig:LIGO-Detections-1}
\end{figure}
\begin{figure}[t]
	\centering
	\includegraphics[width = 0.45 \textwidth]{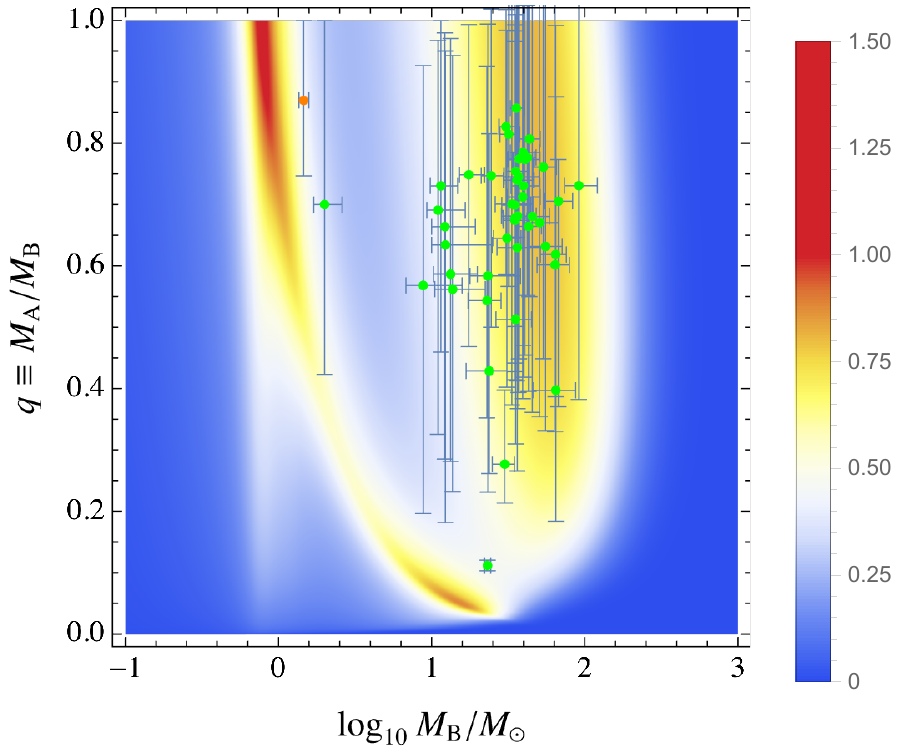}
	\vs{-1.5mm}
	\caption{Same as Fig.~\ref{fig:LIGO-Detections-1} 
			but for case ({\it ii}$\mspace{1.5mu}$)
			with $\ell_{e} = 0$ and $\ell_{\mu} = -\,\ell_{\tau} = 4 \times 10^{-2}$.
			\vs{-4mm}
			}
	\label{fig:LIGO-Detections-2}
\end{figure}

\begin{figure}
	\centering
	\includegraphics[width = 0.45 \textwidth]{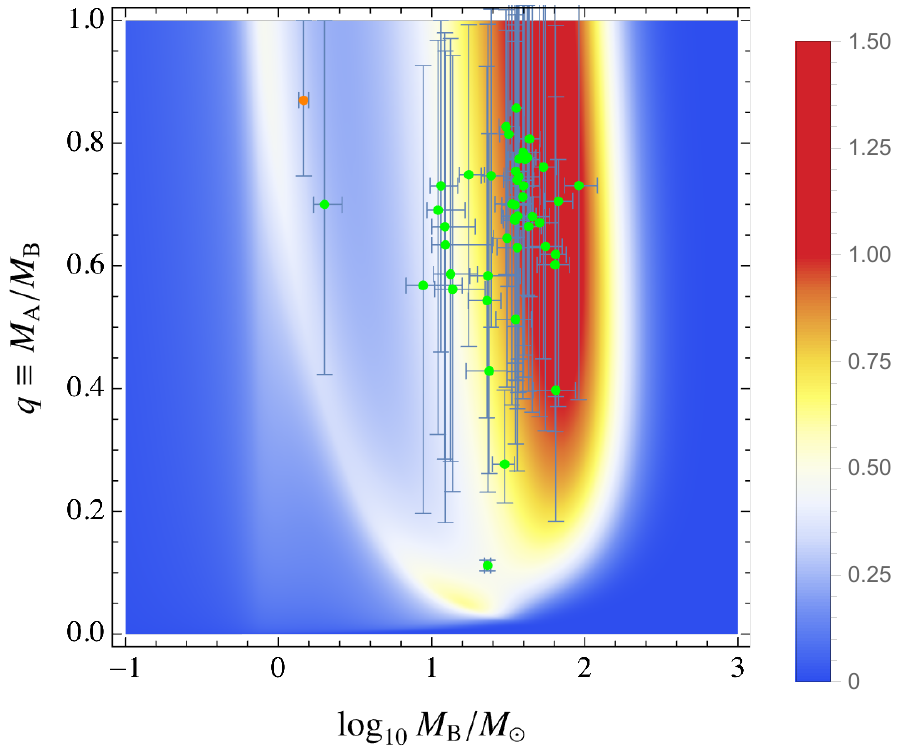}
	\caption{Same as Fig.~\ref{fig:LIGO-Detections-1} 
			but for case ({\it iii}$\mspace{1.5mu}$)
			with $\ell_{e} = -\,8 \times 10^{-2}$ and $\ell_{\mu} = \ell_{\tau} = 4 \times 10^{-2}$.
			}
	\label{fig:LIGO-Detections-3}
\end{figure}

\section{Predicted Merger Rates}
\label{sec:Predicted-Merge-Rates}

The expected distribution of PBH mergers which can be detected by LIGO/Virgo can be estimated for binaries formed through tidal capture in dark-matter halos. Using the merger rate $\tau$ of two merging black holes with masses $M_{\rm A}$ and $M_{\rm B}$ as given in Ref.~\cite{Clesse:2016vqa} (\cf~Refs.~\cite{Raidal:2018bbj, DeLuca:2020qqa} and references therein) and the detector range $R$ from Ref.~\cite{Chen:2017wpg} utilising the sensitivity curve from Ref.~\cite{LIGOScientific:2018mvr}, the event likelihood $E$ scales as $E \propto \tau\.R^{3}$ (cf.~Ref.~\cite{Carr:2019kxo}). Note that the detector range $R$ involves an integral over frequency, from the minimum frequency seen by the detector up to the merger frequency, or the maximal frequency the detector is able to observe (as far as light binaries are concerned) (see Ref.~\cite{Chen:2017wpg}). Figures~\ref{fig:LIGO-Detections-1} to \ref{fig:LIGO-Detections-3} visualise the likelihood $E$ as a function of the larger PBH mass $M_{\Brm} > M_{\Arm}$ and the mass ratio $q \equiv M_{\Arm} / M_{\Brm}$ for the three cases ({\it i}$\mspace{1.5mu}$) -- ({\it iii}$\mspace{1.5mu}$). Additionally we show all currently known black-hole mergers (green dots) as reported by the LIGO/Virgo collaborations \cite{Abbott:2020gyp}. The lepton flavour symmetric case predicts overly many events for small masses and/or small $q$ (see Fig.~\ref{fig:LIGO-Detections-1}) due to the pronounced peak at the QCD transition. On the contrary, both cases of unequal lepton flavour asymmetries  (Figs.~\ref{fig:LIGO-Detections-2}, \ref{fig:LIGO-Detections-3}) yield an enhanced likelihood for merges of PBHs for masses of order $100\,M_{\odot}$, where most of the events are concentrated.

\section{Conclusion}
\label{sec:Conclusion}

We have pointed out that lepton flavour asymmetries{\;---\;}within the boundaries allowed by current observational constraints{\;---\;}has a pronounced effect on the PBH formation rate. Most dominantly, it always lowers the peak in the PBH mass function around the QCD transition, thereby giving more relative weight to the pion/muon plateau at a few ten to around hundred solar masses. If PBHs are produced during the pion and muon annihilation epoch and with a substantial dark-matter fraction, lepton flavour asymmetries significantly alter the prediction for the PBH mass spectrum and therefore their mergers, which might be observed with LIGO/Virgo. Together with an improved understanding of the BH mass spectrum from core collapse events and their corresponding BH merger rates, upcoming gravitational-wave data will thus yield a new way to probe how lepton flavour asymmetric the Universe has been prior to the onset of neutrino oscillations.

\acknowledgements
{\it Acknowledgements\;---\;}We are grateful to Valerio De Luca, Gabriele Franciolini, Antonio Riotto, J{\"u}rgen Schaffner-Bielich, York Schr{\"o}der and Hardi Veerm{\"a}e for valuable remarks. F.K.~thanks Bernard Carr, S{\'e}bastien Clesse and Juan Garc{\'i}a-Bellido for stimulating discussions and helpful input from the fruitful collaboration on the work of Ref.~\cite{Carr:2019kxo}, which builds a cornerstone for this work. Support by S{\'e}bastien Clesse regarding the {\it Mathematica} code for the expected merger distribution used for Figs.~\ref{fig:LIGO-Detections-1}-\ref{fig:LIGO-Detections-3} is particularly acknowledged. D.B.~and D.J.S.~acknowledge support by the Deutsche Forschungsgemeinschaft (DFG) through the Grant No.~CRC-TR 211 ``Strong-interaction matter under extreme conditions''. I.M.O.~acknowledges support from FPA2017-845438 and the Generalitat Valenciana under grant PROMETEOII/2017/033.

\bibliography{refs}

\end{document}